\documentclass[11pt]{aastex}
\usepackage{graphicx}
\usepackage{amsmath} 
\usepackage{natbib}

\usepackage{emulateapj5}
\usepackage{onecolfloat}

\newcommand{\beqa}{\begin{eqnarray}}
\newcommand{\eeqa}{\end{eqnarray}}
\newcommand{\nn}{\nonumber}

\newcommand{\UnitMpc}{\:\mathrm{Mpc}}
\newcommand{\Unitkpc}{\:\mathrm{kpc}}
\newcommand{\UnitMsun}{\:M_\odot}
\newcommand{\Unitkmpers}{\:\mathrm{km/s}}

\def\hMpc{\ifmmode{h^{-1}{\rm Mpc}}\else{$h^{-1}{\rm Mpc}$}\fi}
\def\hkpc{\ifmmode{h^{-1}{\rm kpc}}\else{$h^{-1}{\rm kpc}$}\fi}
\def\hMsun{\ifmmode{h^{-1}M_\odot}\else{$h^{-1}M_\odot$}\fi}

\newcommand{\CTR}[1]{\multicolumn{1}{c}{#1}}

\begin{document}

\title{
	Velocity dispersion profile in dark matter halos}

\shorttitle{Dispersion in dark matter halos}
\shortauthors{Hoeft et al.}

\author{M. Hoeft \altaffilmark{3}}
\affil{International University Bremen \altaffilmark{1}}
\author{J.P. M\"ucket \altaffilmark{4}} 
\author{S. Gottl\"ober \altaffilmark{5}}
\affil{Astrophysikalisches Institut Potsdam\altaffilmark{2}}

\altaffiltext{1}{Campusring 1, 28759 Bremen, Germany }
\altaffiltext{2}{An der Sternwarte 16, 14482  Potsdam, Germany }
\altaffiltext{3}{m.hoeft@iu-bremen.de}
\altaffiltext{4}{jpmuecket@aip.de}
\altaffiltext{4}{sgottloeber@aip.de}

\begin{abstract}

Numerous numerical studies indicate that dark matter halos show an almost 
universal radial density profile. The origin of the profile is still under 
debate. We investigate this topic and pay particular attention to the velocity 
dispersion profile. 
To this end we have performed high-resolution simulations
with two 
independent codes, ART and {\sc Gadget}. The radial velocity dispersion can be  
approximated as function of the potential by $\sigma_r^2 = a (\Phi / 
\Phi_{\rm{out}} )^\kappa ( \Phi_{\rm{out}} - \Phi )$, where $\Phi_{\rm{out}}$ is 
the outer potential of the halo.
For the parameters $a$ and $\kappa$ we find
$a=0.29\pm0.04$ and $\kappa=0.41\pm0.03$. 
We find that the power-law asymptote $\sigma^2 \propto 
\Phi^\kappa$ is valid out to much larger distances from the 
halo center than any power asymptote for the density profile $\rho \propto r^{-n}$. 
The asymptotic slope $n(r \to 0 )$ of the density profile is related 
to the exponent $\kappa$ via $n=2\kappa/(1+\kappa)$. Thus the value obtained for 
$\kappa$ from the available simulation data can be used to obtain an estimate 
of the density profile below presently resolved scales. 
We predict a continuously decreasing $n$ towards the halo center with
the asymptotic value $n \lesssim 0.58$
at $r=0$.

\end{abstract}

\keywords{
	cosmology:theory,
	dark matter,
	galaxies: formation,
	galaxies: structure,
	methods: analytical,
	methods: numerical }

\twocolumn

\section{Introduction}

The formation of structures by purely gravitationally interacting cold dark matter 
(CDM) is one of the fundamental paradigms in cosmology.  The luminous baryonic 
matter is embedded in dark matter halos.  Numerical studies of structure 
formation, which allow only for gravitational interactions, predict the 
distribution of galaxies and clusters of galaxies in excellent agreement with 
observations.  Earlier studies had indicated that not only the distribution of halos 
but also their density profiles depend on the underlying cosmological model 
\citep{quinn:86,frenk:88,dubinski:91,crone:94}. However, subsequent numerical 
investigations revealed that the halo density profiles are almost universal  
and depend neither on the mass of the halo \citep{navarro:96,navarro:97}, the 
initial fluctuation spectrum, nor the underlying cosmological model. This 
similarity of dark matter halos is complemented by the discovery of a universal 
angular momentum profile \citep{bullock:01}.

With the dramatically increased resolution of simulations the properties of the 
profiles can now be studied over several orders of magnitude in radius. 
This allows a more detailed study of 
the central slope of the density profile. \citet{navarro:96,navarro:97} 
first gave an analytic  approximation  for the density profiles. They obtained 
$\rho \propto r^{-n}$, where the slope depends on the radius according to $n(r) 
= (1+3r/r_s)/(1+r/r_s)$, $r_s$ denotes the scale radius of a given halo. This 
NFW-profile implies an asymptotic slope of $n = 1$ for the central density 
profile. Other numerical studies produced a significantly steeper inner slope, 
$n \approx 1.5$, 
\citep{moore:98,moore:99,ghigna:98,ghigna:00,fukushige:01,klypin:01}.  More 
recently \citet{power:02} found no asymptotic slope at all but a continuously 
decreasing slope with values down to $\approx 1.2$ approaching the innermost 
radius
resolved. In order to determine the inner slope of dark matter halos in 
numerical studies more reliably, more efforts are needed to increase the 
resolution of simulations.

In galaxies the motion of the stars is governed by the dark matter halo. 
Measuring rotation curves in the very center of a galaxy allows in principle 
the determination of the central profile of the underlying density distribution. 
Low surface brightness galaxies are almost unbiased by any gas and are thus 
promising candidates for the determination of the innermost density profile. A 
`cuspy', 
$n \gtrsim 1$, dark matter core is ruled out by many studies 
\citep{gaugh:98,blok:01} including models of the inner rotation 
curve of the Milky Way \citep{binney:01}. 
\citet{salucci:00} have inferred a constant central 
density from the kinematics of a large 
sample of spiral galaxies. However, some recent investigations of dwarf galaxies 
suggest that an innermost slope of $n \approx 1$ is consistent with rotation 
curve data \citep{bosch:01}.  Lensing studies of clusters of galaxies indicate that 
there is a shallow core, with some preference of an isothermal profile in the very 
center \citep{seitz:98,gavazzi:03}. Thus, present observations 
favor an inner slope of $n \approx 1$ at most and in many cases a considerably 
shallower one.

Some mechanisms have been proposed which are able to erase a central cusp of dark 
matter halos.  \citet{el-zant:01}, e.g., suggest that a core may be eliminated 
by dynamical friction of an initially clumpy gas distribution. Alternatively the 
occurrence of bar instabilities in the center have been discussed 
\citep{weinberg:02,sellwood:03}. \citet{dekel:03} have argued that a central cusp can 
be avoided by puffing up infalling halos, which could occur due to baryonic 
feedback processes.

The uncertainty of the central slopes reflects that up to now dark matter halo profiles 
have only been obtained on a empirical basis. Efforts are made to calculate 
the profiles analytically and semi-numerical models arew used to calculate the profile 
using the growth rate of halos. Since the mass accretion depends on the 
environment those models may indicate that the profile should also vary with the 
supposed cosmology \citep{syer:98,nusser:99,lokas:00}. An alternative 
explanation was given by \citet{taylor:01,navarro:01}. They argued that  
recurrent merging results in a phase-space density profile which 
decreases according to a power-law. As a result they found that the central 
slope should be as low as $n=0.75$. One may summarize the results about dark 
matter density profiles by stating the open questions. Why do halos show a profile 
close to the NFW-profile, why are the profiles almost universal, and what is 
the central slope?

In this paper we investigate the velocity dispersion profile of dark matter 
halos that were simulated with high resolution and using two independent codes. In 
Sec.~\ref{sec-vdpr} we present a suitable formula that approximates 
the dispersion profile as a function of the potential. The probability 
distribution of the particle velocities within shells of different 
radii is investigated in Sec.~\ref{sec-pdf}. This allows us to interpret 
the individual factors which are present in the suggested formula. The analysis 
of the possible asymptotic behavior of the profiles in Sec.~\ref{sec-asymp} demonstrates 
that the relation between velocity dispersion and potential is strongly 
restricted by physical requirements. Moreover, limits for the inner slope of the 
density profile can be set. For consistency we show in 
Sec.~\ref{sec-jeans} that the numerical density profiles can be recoverd by 
integrating the basic equations.

\begin{figure}[t]
	\centering
	\resizebox{1.02\hsize}{!}{\includegraphics[angle=0]{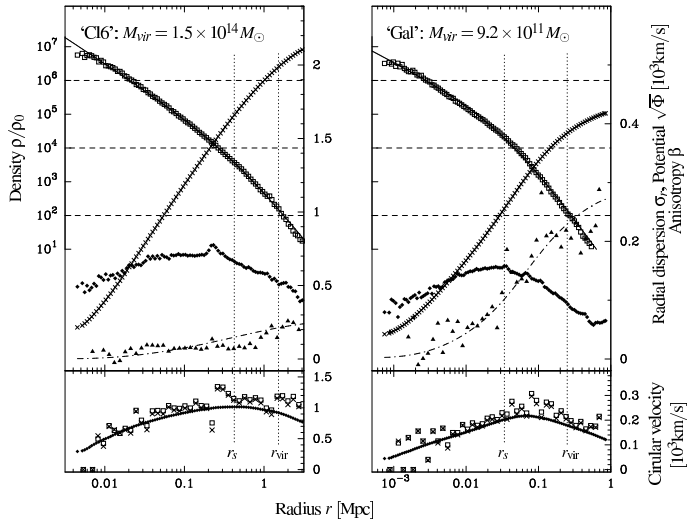}}
   \caption{ The radial profiles with respect to different quantities
   for two halos (galaxy sized and cluster sized).  Upper panel:
   The density $\rho/\rho_0$ (open squares, scaling given by left-hand y-axis), 
the potential
   $\sqrt{\Phi}$ (crosses, scaling by right-hand y-axis), the radial velocity 
dispersion $\sigma_r$
   (solid squares, scaling by right-hand y-axis) and the anisotropy $\beta$ 
(triangles, scaling by right-hand y-axis) are 
plotted.
	Potential and dispersion are given in units of $[10^3\Unitkmpers]$ 
	what allows us to compare immediately these quantities to the 
	velocity p.d.f. Fig.~\ref{fig-ehist}.
   The density is approximated by generalized NFW-profile
   Eq.~(\ref{eq-nfw-profile}) (solid line) with parameters $r_s = 0.42
   \UnitMpc$, $\delta_c = 1.01 \times 10^4$, $n = 1.54$ and $r_s =
   0.034 \UnitMpc$, $\delta_c = 7.1 \times 10^4$, $n = 1.11$ for
   the cluster-sized halo Cl6 and the galaxy-sized halo Gal, respectively. 
  The anisotropy is approximated by $\beta = 0.27 * \Phi / \Phi_{\rm{out}}$. The 
values of $\Phi_{\rm{out}}$ are taken from Tab.~\ref{table-halos}. 	
	Lower panel: Predictions by the left hand side of the Jeans equation
   (\ref{eq-Jeans}) including the anisotropy term (crosses) and
   excluding it (open squares) match the circular velocity (solid
   squares).  
	}
	\label{fig-profiles}
\end{figure}

\section{Radial profiles from high-resolution simulations}

\label{sec-simu}

We have performed several dark matter high-resolution simulations. We have 
assumed a spatially flat cold dark matter model 
with a cosmological constant favored by most current observations.  For the 
first set of simulations we have used the Adaptive Refinement Tree (ART) 
$N$-body code \citep{kravtsov:97} and the following cosmological parameters: 
$\Lambda$CDM with $\Omega_{\rm M}=0.3$, $\Omega_{\Lambda}=0.7$, $\sigma_8=0.9$, 
and $h=0.7$.  Within a low mass resolution simulation ($128^3$ particles in a 
$114\UnitMpc$ 
cubed 
simulation box, $m_{\rm{part}} = 2.9 \times 10^{10}\UnitMsun$), we 
have identified clusters of galaxies. From this set we have selected 8 
candidates with different masses and merging histories and added 5 smaller 
clusters/groups for numerical load balance. We then re-simulated the clusters 
with higher mass resolution. With particle masses of $4.6 \times 
10^{8}\UnitMsun$ a typical cluster and its environment contains more than one 
million particles. The highest force resolution with 9 refinement levels was 
$0.9 \Unitkpc$. Subhalos with masses above $4.6 \times 10^{10}\UnitMsun$ are 
well resolved. A typical cluster contains more than 150 such subhalos. The 
simulations were done using an MPI version of the ART code where each of eight 
nodes 
followed the evolution of one or two clusters. Another simulation within a box 
of $36 \UnitMpc$ 
box length 
contains a galaxy-sized halo. The region containing this halo 
was simulated with an 
effective resolution of
$1024^3$ particles, i.e. with a mass resolution of 
$1.7 \times 10^6 \UnitMsun$. The highest force resolution with 10 refinement 
levels was $0.1 \Unitkpc$.  

In addition, we investigate the properties of a cluster-sized halo
obtained by a high-resolution simulation with uniform mass for all
particles. This halo is resolved by more than one million
particles since a huge entire number of particles and a comparatively
small simulation box was used, namely $300^3$ particles and a box size
of $30 \UnitMpc$. The initial conditions were set up according to the
cosmological model: $h=0.65$, $\Omega_{\rm M}=0.3$, and
$\Omega_{\Lambda}=0.7$.  The simulation has been performed using the
public {\sc Gadget}-code \citep{springel:01}. For comparison we consider another
simulation, performed with the {\sc Gadget}-code, with the same initial
conditions as the cluster-sized halo Cl6 simulated with ART. See
Tab.~\ref{table-halos} for a compilation of the halo properties.

We have determined the radial density profiles for all considered objects at 
redshift $z = 0$.  In Fig.~\ref{fig-profiles} the density profiles of a 
cluster-sized and a galaxy-sized halos are shown. Both profiles can be fitted 
reasonably well by a generalized NFW-profile with a free parameter $n$ for 
the inner slope
\beqa
\rho/\rho_0 
	&=&
  	\delta_c (r/r_s)^{-n} (1+r/r_s)^{-(3-n)}
	.
	\label{eq-nfw-profile}
\eeqa	
The parameters $n$, $\delta_c$, and $r_s$ are determined for each halo by a 
least-square 
fit to the mean densities in radial bins up to the virial radii. For the 
halo Cl6 we obtain an inner slope of $n \approx 1.54$, which corresponds 
to the Moore-profile and which is in agreement with the results given by 
\citet{fukushige:01}. They analyzed 12 halos with various masses and found 
an inner slope of about $n \approx 1.5$ for all of them. In contrast, 
for most clusters of our sample we found smaller inner slopes, see 
Tab.~\ref{table-halos}.

A relaxed spherical halo of collisionless particles is completely described by 
the 
radial profiles of the density $\rho(r)$, the radial velocity dispersion 
$\sigma_r^2 = \overline{(v_r-\overline{v_r})^2}$, where $\overline{v_r}$ is the 
mean radial velocity in a spherical shell with mean radius $r$, and the 
anisotropy of the dispersion $\beta = 1- \sigma_t^2 /\sigma_r^2$, where 
$\sigma_t$ denotes the tangential velocity dispersion. The potential can be 
obtained by mass integration $\Phi = G\, \int^r_0 \, dr M(r) / r^2$ (Poisson 
equation), where $G$ denotes the gravitational constant. The radial profiles are 
related by the Jeans equation \citep{binney:87}
\beqa
\frac{d ( \rho \sigma^2_r )} {dr}
	+ \frac{2\rho}{r} \,\beta \sigma_r^2  
	&=&
	- \rho \, \frac{d \Phi}{dr}
	,
	\label{eq-Jeans}
\eeqa	
which describes a steady-state halo whose particles move on collisionless  
trajectories in a spherical potential self-consistently generated by the 
particle distribution. 
The results of numerical simulations confirm that dark matter halos 
fulfill the Jeans equation at least up to the virial radius \citep{thomas:98}. 
This leads to the conclusions that (i) the halos are  relaxed, (ii) the 
particles in the halos are moving on orbits determined by a mean potential, 
and (iii) the small-scale particle-particle interaction is negligible, i.e. 
close encounters are rare.  

We have tested whether our halos fulfill the Jeans equation.  The circular 
velocities $v_c$ are defined by $v_c^2= G M(r) / r = r \Phi' $. Using the Jeans 
equation they can also be given by $-r(\rho \sigma_r^2)'/\rho - 2 \beta 
\sigma_r^2$, where the prime denotes $d/dr$.  In Fig.~\ref{fig-profiles}, lower 
panels, we compare the circular velocities obtained by both ways.  Furthermore, 
we compute the circular velocities neglecting any anisotropy contribution 
$(\beta = 0)$.  The differences between the circular velocities at 
given radius are small and, hence, the halos can approximately be described by 
the Jeans equation assuming an isotropic velocity dispersion.

\begin{figure}[t]
	\centering
	
\resizebox{0.7\hsize}{!}{\includegraphics[angle=0]{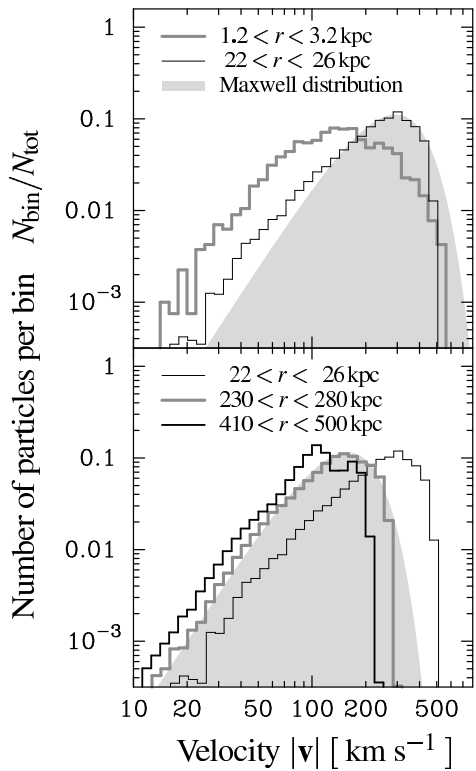}}
	\caption{
	The distribution of the absolute value of the particle
	velocities $|\mathbf{v}|$ within spherical shells 
	for the galaxy-sized halo Gal.
The distributions shown in the upper panel are taken in the shell around the 
radius $r_{\sigma,{\rm max}}$ with maximum dispersion and in a second shell with 
smaller radius. The distributions shown in the lower panel are taken in shells 
starting from $r_{\sigma,{\rm max}}$ to larger radii. 
	For comparison, a Maxwell distribution is shown 
	centered at the maximum of the velocity distribution within 
	one of the shells.
}
\label{fig-ehist}
\end{figure}

\section{Velocity dispersion as a function of the potential} 

\label{sec-vdpr}

A spherically symmetric halo with isotropic velocity dispersion can be described 
using three radial functions, e.g. density, potential and velocity dispersion. 
These profiles are
related by the Jeans and the Poisson equation. Thus, if one profile 
or any supplementary relation between $\sigma, \rho$ or $\Phi$ 
is given, the two others are determined. An invitingly simple way to get 
the profiles is to assume that the velocity dispersion is constant within the 
virial radius. 
However, the numerical results indicate 
that the velocity dispersion has a pronounced maximum at a radius 
$r=r_{\sigma,\rm{max}}$. For example, the galaxy-sized halo Gal shows 
the maximum velocity dispersion of $155\Unitkmpers$ at the 
position $r \approx 30 \Unitkpc$. In contrast the dispersion is only $90\Unitkmpers$ at 
the virial radius and to $80\Unitkmpers$ at the innermost ($r\approx 1\Unitkpc$) 
radius.

\begin{figure}[t]
	\centering
	\resizebox{0.9\hsize}{!}{\includegraphics[angle=0]{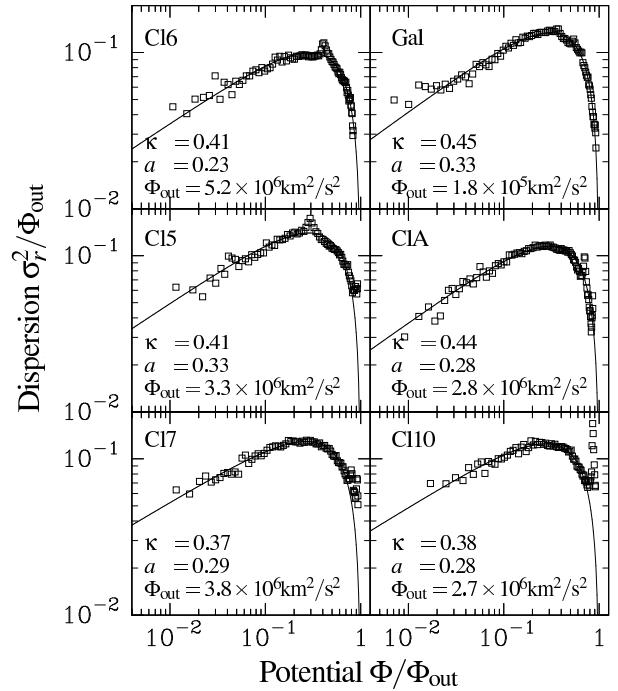}}
	\caption{ 
	The dependence of the radial velocity dispersion
	$\sigma_r^2/\Phi_{\rm{out}}$ on the potential $\Phi/\Phi_{\rm{out}}$.  
	The data obtained from the simulations are approximated by 
	the VDPR~(\ref{eq-of-state}). 
	} 
	\label{fig-sig-phi}
\end{figure}

On the basis of the numerical results we make an heuristic ansatz for 
the relation between the velocity dispersion and the potential at a given radius. 
Physical arguments which lead to this ansatz are discussed in Sec.~\ref{sec-pdf}.
We approximate the dispersion as a function of the potential. This can be done 
because the potential increases monotonically with radius and, consequently, a 
one-to-one mapping between radius and potential must exist.  The data can be 
approximated, see Fig.~\ref{fig-sig-phi}, by the relation 
\beqa
	\sigma_r^2
	&=&
	a
	\left( \frac{ \Phi } { \Phi_{\rm{out}} }  \right)^\kappa
	\left( \Phi_{\rm{out}} - \Phi \right)
	,
	\label{eq-of-state}
\eeqa	
where $a$, $\kappa$ are free parameters and $\Phi_{\rm{out}}$ is the maximum 
potential reached for large radii. We denote this relation between radial 
velocity dispersion and potential in the following as VDPR. Since an arbitrary 
constant value can always be added to the potential we have set $\Phi(0)=0$. 
This normalization of the potential is used throughout this paper. We perform 
least-square fits to the data of the different halos up to radii of about two 
times the virial radius. Some of the halos have a companion or substructures 
which results in a narrow peak in the dispersion profile, e.g. halo Cl10. 
We exclude those peaks from the approximation if they are beyond the virial 
radius. The mean values obtained for the dimensionless parameters are $a = 
0.29\pm 0.04$ and $\kappa = 0.41 \pm 0.03$, where the errors are standard 
deviations.

The primary halo-specific parameter is the outer potential  $\Phi_{\rm{out}}$. 
Also the parameters  $a$ and $\kappa$ vary, but their scatter is small. It may 
be caused by substructures still present in the halo, imperfect relaxation, or 
even nearby structures which disturb the spherical symmetry. The smallness of 
the scatter may indicate that perfectly relaxed halos have only one free 
parameter, namely the outer potential. Consequently the VDPR may be considered 
to be equivalent to an equation-of-state since it gives the velocity dispersion 
as a function of the local potential, independent of the radius. On the other 
hand it incorporates 
global properties, namely the difference of the local potential to the central 
and to the outer potential. This reflects that the local particle-particle 
interaction is small and particles do not significantly exchange energy. 
Therefore, the local velocity dispersion must depend on the global properties. 
However, if for all perfectly relaxed halos both parameters $a$ and $\kappa$ are the same,
the VDPR may 
serve as an additional equation which allows to close the system of Jeans and 
Poisson equation.

\section{Velocity distribution in spherical shells}

\label{sec-pdf}

The VDPR consists apparently of two parts. The first factor $( \Phi / 
\Phi_{\rm{out}}  )^\kappa$ is dominant in the center of the halo whereas the 
second factor $\left( \Phi_{\rm{out}} - \Phi \right)$ is dominant in the outer 
region. Also the shape of the velocity probability density function (p.d.f.) 
in a shell with a small radius differs significantly from the shape in a 
shell with a large radius.

In Fig.~\ref{fig-ehist} we provide the velocity p.d.f. 
determined within several spherical 
shells of the galaxy-sized halo Gal. Let us first consider the velocity p.d.f. 
$h(|\mathbf{v}|,r)$ in the upper panel in a shell around $ 22<r< 26 \Unitkpc$ 
close to the radius $r_{\rm{\sigma , max}}$, where the velocity dispersion 
reaches its maximum. In comparison to a Maxwellian distribution the velocity 
p.d.f.  $h(|\mathbf{v}|,r_{\rm{\sigma\,max}})$ is flatter at the 
small-velocity-tail and shows a steep break towards high velocities.  The second 
p.d.f. for the shell shown in the upper panel has a radius much smaller than 
$r_{\rm{\sigma , max}}$.  The velocity p.d.f. 
is 
much broader compared with a Maxwellian distribution.  The probability peaks at 
$\approx 100 \Unitkmpers$. Albeit there are also particles with velocities 
$\gtrsim 300 \Unitkmpers$ -- but below the escape velocity -- the probability to 
find a particle with such a large velocity is very small near the center of the 
halo. This effect could be due to two reasons: first, the angular momentum of a 
high velocity particle must be very small to get close to the center and second, 
such a particle spends only very little time along its orbit near the center. 
Thus the probability to find particles with a high velocity decreases towards 
the center of the halo, whereas the orbit of slow particles is entirely in the 
vicinity of the center. Thus the potential well acts to concentrate slow particles in 
the central regions. Since there is no characteristic radius for this 
concentration process one may expect that the velocity dispersion scales at small radii 
in a self-similar manner with the potential, i.e. $\sigma_r^2 \propto  
\Phi^\kappa$. This is represented by the first factor of the VDPR.

Let us now consider the shape of the velocity p.d.f. within shells around and 
above the virial radius, see Fig.~\ref{fig-ehist}, lower panel.  With increasing 
radii of the shells the entire distribution shifts to smaller velocities. The 
shape of the velocity p.d.f. itself almost does not change. The probability 
distribution for small velocities, i.e. below the maximum of the distribution 
$v_{\rm{max}}$,  converges to the Maxwellian distribution. For larger velocities 
the steep break is still present.  An obvious reason for the shift is that the 
kinetic energy decreases when a particle moves out of the potential well. 
This implies that the characteristics of the velocity p.d.f., e.g. the position 
of the maximum velocity $v_{\rm{max}}(r)$ and the velocity dispersion 
$\sigma(r)$, depend directly on the potential $\Phi(r)$. Particles having 
velocities different from zero for $r \to \infty$, would leave the halo, i.e. 
they are not bound. Consequently the dispersion should vanish for large 
radii for an isolated halo with no additional matter around. Thus, 
interpreting the shift of the velocity p.d.f. as a result of the effective 
potential well and assuming that the shape of the p.d.f. is constant for all 
radii suggests a dependence between dispersion and potential according to 
$\sigma^2(r) \propto (\Phi_{\rm{out}} - \Phi(r))$ at large radii. This is 
represented by the second factor in the VDPR.

\begin{center}
\begin{table*}
\begin{center}
\begin{tabular}{lcrrrrrrrr}
label & Code & 
	\CTR{$r_{\rm{virial}}$}  & 
	\CTR{$M_{\rm{virial}}$}  & 
	\CTR{$N_{\rm{virial}}$}  & 
	\CTR{$c$}  &
  	\CTR{$n$}   & 
	\CTR{$\Phi_{\rm{out}}$}  & 
	\CTR{$\kappa$} & 
	\CTR{$a$} \\
&
	&
	\CTR{$[{\UnitMpc}]$}  &   
	\CTR{$[{\UnitMsun}]$} &
	&&& 
	\CTR{$[\rm{km^2/s^2}]$} 
	&& \\
\hline
Gal & 
	ART &
	0.25 & $9.2\times10^{11}$ & 1066000 &
	7.4 & 1.11 & 
	$1.8\times10^5$  &  0.45 & 0.33 \\
ClA &
	Gad &
	0.90 & $7.0\times10^{13}$ & 1031000 &
	8.8 & 1.17 &
	$2.8\times10^6$  &  0.44 & 0.28 \\
Cl6 & 
	ART &
	1.52 & $1.46\times10^{14}$ & 646000 &
	3.6 & 1.54 & 
	$5.2\times10^6$  &  0.41 & 0.23 \\
Cl6b & 
	Gad &
	1.30 & $1.34\times10^{14}$ & 590000 &
	3.6 & 1.47 & 
	$5.2\times10^6$  &  0.41 & 0.24 \\
Cl7 & 
	ART &
	1.27 & $1.23\times10^{14}$ & 541000 &
	5.2 & 1.14 & 
	$3.8\times10^6$  &  0.37 & 0.29 \\	
Cl5 & 
	ART &
	1.19 & $1.02\times10^{14}$ & 452000 &
	5.0 & 1.12 & 
	$3.3\times10^6$  &  0.41 & 0.33 \\	
Cl10 & 
	ART &
	0.97 & $5.59\times10^{13}$ & 247000 &
	5.7 & 1.27 & 
	$2.7\times10^6$  &  0.38 & 0.28 \\	
Cl9 & 
	ART &
	0.90 & $4.38\times10^{13}$ & 193000 &
	8.9 & 0.99 & 
	$2.4\times10^6$  &  0.47 & 0.34 \\		
Cl12 & 
	ART &
	0.77 & $2.79\times10^{13}$ & 123000 &
	3.5 & 1.46 & 
	$1.6\times10^6$  &  0.40 & 0.29 \\	
\hline
Mean & & & & & & & &
	0.41 & 0.29 \\
\end{tabular}
\end{center}
	\caption{ \rm
	Parameters of the halos obtained by high resolution $N$-body 
simulations.	
	The virial radius is determined according to 
	$M_{\rm{vir}} \: /  \:(3/4) \pi r_{\rm{vir}}^3 
	= 178 \, \Omega_0^{0.45} \rho_{\rm{crit}}$, where $\rho_{\rm{crit}}$ 
denotes the
	critical density for a flat universe, see \cite{eke:98} 
	The density profile of the halos is approximated using a 
	least-square fit to the (logarithmic) density values by a 
	generalized NFW-profile Eq.~(\ref{eq-nfw-profile}). 
	The concentration is given by $c = r_{\rm{vir}}/r_s$.
	The relation between the radial velocity dispersion $\sigma_r^2$ and the
	gravitational potential is approximated by the VDPR~(\ref{eq-of-state}) 
	using a least-square fits 
	with the free parameters $a$, $\kappa$ and $\Phi_{\rm{out}}$.
 	}
	\label{table-halos}
\end{table*}
\end{center}

\section{Asymptotic profiles}

\label{sec-asymp}

We now consider the asymptotic behavior of the radial profiles at small, $r \ll r_s$, and at 
large radii, $r \gg r_s$.
	
First we discuss the outer region, where the dispersion can be approximated by 
$\sigma_r^2 |_{r \gg r_s} = a \,(\Phi_{\rm{out}} - 
\Phi(r))$. Using the Jeans equation allows we can relate the factor 
$a$ to the slope of density profile. We assume isotropic dispersion, substitute 
the radial dispersion in the Jeans equation by the 
potential, integrate the Jeans equation and finally obtain the relation  
\beqa
	\left. \rho(\Phi) \, \right|_{r \gg r_s} 
	&=&
  	c \: (\Phi_{\rm{out}} - \Phi)^\gamma, 
	\qquad 
	\gamma = (1-a)/a
	.
	\label{eq-rho-phi-isotrop}
\eeqa
Since, both,  $(\Phi_{\rm{out}} - \Phi)$ and the density $\rho$, 
are monotonically decreasing functions for dark matter 
halos, the exponent must 
be positive. 
Therefore, the factor $a$ has to 
be in the range $0<a<1$. 
We can constrain $a$ further by inserting 
Eq.~(\ref{eq-rho-phi-isotrop}) into the Poisson equation. This results in the 
Lane-Emden equation \citep{binney:87}
\beqa
	\frac{1}{r^2}
	\frac{\rm{d}}{\rm{d}r}
	\left\{
			r^2 \frac{\rm{d}}{\rm{d} r }(\Phi_{\rm{out}} -\Phi)
	\right\}
	+
	4 \pi G c (\Phi_{\rm{out}} -\Phi)^\gamma
	&=&
	0
	.
	\label{eq-lane-emden}
\eeqa
One solution of the Lane-Emden equation is a power-law for the relative 
potential
\beqa
	(\Phi_{\rm{out}} -\Phi)
	&\propto&
	r^{-m},
	\qquad
	m = 2/(\gamma-1)
	\nn
\eeqa
and with Eq.~(\ref{eq-rho-phi-isotrop}) also for the density
\beqa
	\left. \rho \right|_{r \gg r_s}
	\: \propto \:
	(\Phi_{\rm{out}} -\Phi)^\gamma
	&\propto&
	r^{-n},
	\qquad
	n = 2\gamma/(\gamma-1) = (2-2a)/(1-2a)
	.
	\nn
\eeqa
In order to obtain a solution for a finite mass halo the slope of density profile must 
be sufficiently 
steep, namely $n>3$. 
Therefore, the factor $a$ must be in the range 
\beqa
	1/4<a<1/2
	.
	\label{eq-restrict-a}
\eeqa
Note the asymptotic slope of the NFW-profile, $n(r\to \infty)= 3$, 
results from $a=1/4$. 
The simulations show that the anisotropy of the velocity dispersion  
increases with radius and amounts to $\beta \approx 0.3$ at the virial radius. 
This anisotropy affects the allowed 
parameter range for $a$: Integrating the 
Jeans equation under the condition of a constant anisotropy parameter $\beta$ 
leads to 
\beqa
	\left. \rho(\Phi,r) \, \right|_{r \gg r_s } 
	&=&
	c^\ast \:
	(\Phi_\infty -\Phi)^\gamma
	\:
	r^{-2\beta}
	.
	\label{rho-outer}
\eeqa	
Inserting this relation again into the Poisson equation allows us to find 
power-law solutions for the potential and the density 
\beqa
(\Phi_\infty -\Phi)
	&\propto&
	r^{-m},
	\qquad
	m = 2(1-\beta)/(\gamma-1)
	\nn
	\\
\left. \rho \right|_{r \gg r_s}
	&\propto&
	r^{-n},
	\qquad
	n = 2(\gamma-\beta)/(\gamma-1) = 2(1-a-a\beta)/(1-2a)
	.
	\nn
\label{anisotr}
\eeqa
Therefore, the factor $a$ has to be in the range $1 / ( 4-2\beta) < a < 1/2$.	
Given the positive anisotropy by the simulations, this narrows the possible 
parameter range for $a$. Thus, supposing that the outer asymptotic slope of the 
density profile is restricted by the demand of a finite halo mass 
even for $r\to\infty$, 
the parameter $a$ has to be in the given narrow range. The halos analyzed here
fulfill this condition even if they have to be truncated at a radius where the 
ambient, infalling matter becomes dominant.

Let us now consider the central part of the halos. The slope of the inner 
profile is uncertain  from simulations because of the lack of force resolution 
and also because of the poor particle statistics. Therefore, it is  still an 
open question whether an inner asymptote for the radial density profile exists 
and whether it is possible to determine its power index by numerical 
simulations. We make the weaker assumption that the radial velocity dispersion 
at sufficiently small radii is given by a power-law with respect to the potential  
\beqa
	\left. \sigma_r^2(\Phi)  \, \right|_{r \ll r_s}
   &=&	
  	a \, \Phi^\kappa \,  \Phi_{\rm{out}}^{1-\kappa}
	.
	\label{power}
\eeqa	
In the same manner as described above, we insert the relation (\ref{power}) into 
the Jeans equation, perform the integration and obtain 
\beqa
	\left. \rho (\Phi) \, \right|_{r \ll r_s}
	&=& 
	\rho_0 \;
	\left( \frac{ \Phi }{ \Phi_{\rm{out}} } \right)^{-\kappa} 
	\exp(-\frac{ ( \Phi / \Phi_{\rm{out}} )^{1-\kappa}}{ a(1-\kappa)})
	,
	\label{rho-exponent}
\eeqa
where $\rho_0$ is a free scaling constant. 
For $\kappa>1$ the assumption of a power-law with respect to $r$ for 
$\Phi$ would result in an exponential-like singularity of the density and 
infinite mass at final radii. In order to avoid this, the 
exponent $\kappa$ has to be within the range
\beqa
	0 < \kappa < 1
	.
\eeqa	
Restricting $\kappa$ in this way the exponential term goes to unity at $\Phi\to 
0$ at sufficient small radii, i.e. $( \Phi / \Phi_{\rm{out}} )^{1-\kappa} \ll 
a(1-\kappa) $  and the inner asymptotic profile for the density is given by  
\beqa
	\left. \rho (\Phi) \, \right|_{r \ll r_s}
	&=& 
	\rho_0 \;
	\left( \Phi / \Phi_{\rm{out}} \right)^{-\kappa}
	.
	\label{rho-central}
\eeqa	

Note, for any $\Phi$ sufficiently small with respect to $\Phi_{\rm out}$ the 
asymptote Eq.~(\ref{power}) has the form of a power-law. In contrast, 
the density approaches the asymptotic behavior not until 
$( \Phi / \Phi_{\rm{out}} )^{1-\kappa} \ll a(1-\kappa) $. 
In particular, if $\kappa$ is allowed to get close to unity, this 
latter condition becomes a very strong restriction and is valid only for very 
little potential values. Thus, the power-law like asymptote for $\sigma_r^2$ can 
be 
adopted much further out from the halo center. Having this in mind 
below we will be able to draw conclusions with respect to the density profile 
near the halo center.
The power-law ansatz $\Phi \propto r^m$ provides a solution for the radial 
profiles. More precisely, by inserting the ansatz into the Poisson equation we can
determine the free constants. For the profiles we obtain
\beqa
	\left.
		\Phi(r)  \,
	\right|_{r \ll r_s}
	&=&
	\Phi_{\rm{out}}
	\;
	\left(
		 r / r_0 
	\right)^{2/(1+\kappa)}
	\label{eq-ini-phi}
	\\
	\left.
		\rho(r) \,
	\right|_{r \ll r_s}
	&=&
	\rho_0
	\;
	\left(
		r / r_0 
	\right)^{-2\kappa/(1+\kappa)}
	\label{eq-ini-rho}
	,
\eeqa	
with
\beqa
	r_0^2
	&=&
	\frac{1}{ 2 \pi G } \;
	\frac{3+\kappa}{(1+\kappa)^2}
	\frac{\Phi_{\rm{out}}}{\rho_0}
	.
\eeqa	
The free, halo-specific parameters for the inner asymptotes are 
$\Phi_{\rm{out}}$ and $r_0$
(or $\rho_0$). In the next section we compute the density profiles by 
integrating the 
basic equations. The integration will be 
performed starting from the small radius $r_0$ up to large radii,
where the value $\Phi_{out}$ will be reached. 
The inner asymptotes given above allows us to calculate the necessary inner 
boundary conditions at $r_0$.

\begin{figure}[t]
	\centering
	\resizebox{0.9\hsize}{!}{\includegraphics[angle=0]{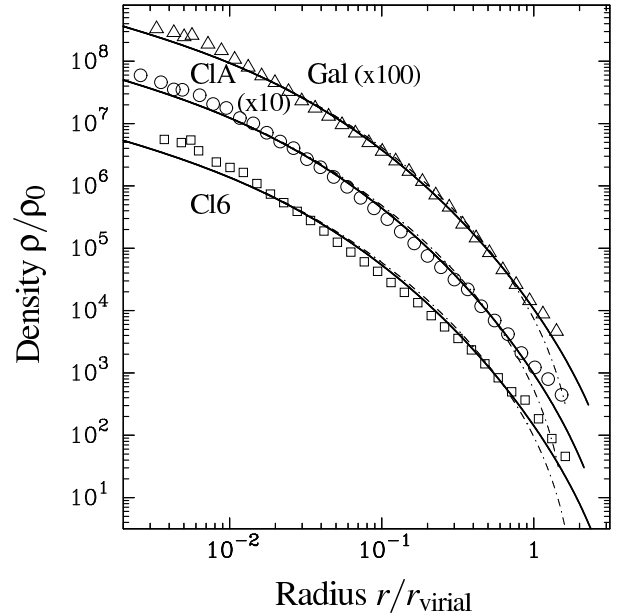}}
	\caption{ Integration of the Jeans equation (\ref{eq-Jeans}).  The
	halos Gal, ClA and Cl6 
	(triangles, circles and squares, respectively) are integrated
	using the VDPR~(\ref{eq-of-state}) with the mean values for $a$ and
	$\kappa$ given in Tab.~\ref{table-halos}. 
  	The profiles are integrated for isotropic velocity dispersion, 
	$\beta = 0$, (dashed line)
	and for an anisotropy according to $\beta = 0.27 \Phi / 
\Phi_{\rm{out}}$, see Fig.~\ref{fig-profiles}. 
	Initial conditions are calculated from Eqs.~(\ref{eq-ini-phi}) and 
	(\ref{eq-ini-rho}). The halo-specific values $\Phi_{\rm{out}}$ are taken
	from Tab.~\ref{table-halos}. For the scale radius $r_0$ we obtained from 
the dispersion   profile $\sigma_r^2(r)$ the values $0.025\UnitMpc$, 
$0.12\UnitMpc$ 
	and $0.13\UnitMpc$ for the halos Gal, ClA and Cl6, respectively.
  	}
	\label{fig-Jeans}
\end{figure}

\section{Integrating the Jeans equation}

\label{sec-jeans}

Using the VDPR we can compute the profiles for a given halo, defined by two 
parameters, e.g.  $\Phi_{\rm{out}}$ and $r_0$. The inner asymptotes are used to 
set the initial conditions. The profiles are obtained by integrating the set 
of ordinary differential equations
\beqa
\frac{d M } {dr}
	&=&
	4 \pi \rho(r) r^2
	\nonumber
	\\
\frac{d \Phi } { dr }
	&=&
	G \: \frac{M}{r^2}
	\nonumber
	\\
\frac{d \sigma_r^2 }{dr }
	&=&
	a \: \frac{d \Phi } { dr } \:
	\left( \frac{ \Phi }{ \Phi_{\rm{out}} } \right)^{\kappa} 
	\left\{ \kappa \frac{\Phi_{\rm{out}}}{\Phi}  - \kappa - 1 \right\}
	\nonumber
	\\
\frac{d \rho } {dr}	
	&=&
	- \frac{ \rho}{ \sigma_r^2 } 
	\left\{
	  		\frac{d \sigma_r^2 }{dr } 
			+ \frac{d \Phi } { dr }
			+ \frac{ 2 \beta \sigma_r^2 } { r }
	\right\}		
	\nonumber
\eeqa
using a Runge-Kutta method. We compare the profiles obtained in this way with 
those from the numerical simulations. The parameter $\Phi_{\rm{out}}$ was 
already calculated by approximating the VDPR, see Tab.~\ref{table-halos}. We 
choose $r_0$ as second free parameter, since it causes only a stretch along the 
radial axis. This parameter can easily be adjusted by approximating the maximum 
position in the dispersion profile $\sigma_r^2(r)$. First, we assume that the 
velocity dispersion is isotropic, $\beta=0$. The resulting density profiles 
approximate roughly the results from the simulations. Clear deviations occur in 
the fringes of the halo. This are the regions where considerable anisotropy in 
the velocity dispersion is present. 
The anisotropy can be reasonably approximated by $\beta = \beta_0 
\Phi / \Phi_{\rm{out}}$ with $\beta_0 \approx 0.27$. After integration, this 
clearly leads to a better concordance of the density profiles, see 
Fig.~\ref{fig-Jeans}.
For the fit of $\beta$ a dependence on the potential $\Phi$ was assumed in 
order to avoid the implementation of additional parameters. It can be easily 
seen that $\beta$ roughly follows the shape of $\Phi$, see 
Fig.~\ref{fig-profiles}. In addition, 
the results are almost insensitive with respect to variations of the dependence 
$\beta$ on $r$ as long as $\beta$ is a monotonic function.

For the halo Cl6, the integration of the differential equations leads still only 
to a poor approximation of the numerical results. This halo shows clear signs 
for undergoing still a merging process. 
The dispersion profile shows a peak at $\approx 1/8 \, r_{\rm{vir}}$, which is 
caused by a remnant of a merger between the cluster and a group about 
$1\;\rm{Gyr}$ ago. Thus this halo is still in the process of relaxation and 
therefore, can be only roughly approximated by
spherical, relaxed system.

The profiles of the halos ClA and Gal are recovered by the 
integration using the mean over all halos for the parameters $a$ and 
$\kappa$. However, all profiles obtained from the simulations are slightly steeper in the center than the integrated curves. \citet{dekel:02} found that the inner profiles of disturbed halos are steepened. 
Therefore, a more cuspy central profile, compared to a perfectly relaxed system, may be caused by some recent merger activity.

The integrated profiles show a sharp break for radii much larger than the virial 
radius. They do not show the asymptotic density profile expected from the 
discussion in Sec.~\ref{sec-asymp}. This is due to the fact that the outer 
potential $\Phi_{\rm{out}}$ is reached at a finite radius, at which the 
dispersion and the density vanishes. However, since the radius where the 
dispersion vanishes is much larger than the virial radius of the 
considered 
halos, this outer slope cannot be determined by simulations designed to 
investigate structure formation.

\section{Discussion and conclusion}

The numerous numerical studies of the structure of dark 
matter halos indicate that perfectly virialized systems show presumably a 
universal 
density profile. The latter can be approximated reasonably well by the 
NFW-profile or 
a generalized version of it. Similarly we have introduced the relation between 
velocity dispersion and potential Eq.~(\ref{eq-of-state}) (VDPR): 
it characterizes the profile by very few parameters. 
In contrast to the approximations of the density profiles  
the VDPR is implicit in the sense that it as a function of the 
potential 
instead of the radius. 
However, by introducing the VDPR we can put restrictions even on the inner 
asymptotic slope of, both, the velocity dispersion profile and the density 
profile.
The data can be approximated very well by the newly introduced VDPR. Almost all 
halos clearly show a power-law for $\sigma_r^2(\Phi) \propto \Phi^\kappa$ 
near the center, the best resolved halos do so over one order of magnitude. 
We suppose that the obtained exponent $\kappa$ reflects also the inner 
asymptotic slope of $\sigma_r^2(\Phi)$. As a result, the innermost density 
profile power index $n$ is related to $\kappa$ by $n=2\kappa/(1+\kappa)$. Using 
the mean value $\kappa = 0.41$, determined from all presented simulations we get 
$n=0.58$. The obtained standard deviation $\Delta \kappa = 0.03$ permits only a 
negligible variation of the exponent $n$. Thus, our prediction for $n$ is  
below the theoretical prediction by \citet{taylor:01}. They 
argued that the inner asymptotic slope is given  by $n=0.75$. Our result also 
indicates that the innermost slope of the density profiles is less steep than 
that obtained from the density profiles directly. This is not a discrepancy since the 
prediction $n=0.58$ is valid for scales smaller than those which are currently 
resolved by numerical simulations. Eq.~(\ref{rho-exponent}) indicates that the 
density profile may steepened by an exponential term in a range where the 
velocity dispersion profile already follows a power-law.  The density profile 
only becomes a power-law if $(\Phi/\Phi_{\rm{out}})^{1-\kappa} \ll a(1-\kappa)$ 
is fulfilled. For instance, the impact of the exponential term becomes smaller 
than 10\% if the potential is below $0.001\times \Phi_{\rm out}$, using the 
obtained values for $\kappa$ and $a$.

In a theoretical analysis of all possible asymptotes \citet{muecket:03} found 
that 
$\kappa$ should be in the range $0 < \kappa < 1/3$. Therefore, the inner 
asymptotic
slope of the density profile should be as low as  $n<0.5$. Our numerical results 
are close to this
theoretical predictions. Possible reasons for the 20\% higher value of $\kappa$ 
could be too strong assumptions 
in the theoretical analysis or steepening of the central density profile during 
relaxation.
In fact, it is still uncertain to which degree dark matter halos are actually 
relaxed. During the cosmological evolution halos merge frequently and show 
almost always substructures. Therefore, the 
present substructure may have impact on the resulting profiles. Nevertheless it 
is assumed that virialization has occured to a high degree and the halo 
configuration 
obeys the Jeans equation. The sample of halos we considered contains very 
relaxed halos, e.g. the halo  Gal, as well as halos with clear deviations from 
relaxation, e.g. the halo Cl6.
\citet{dekel:02} found that ongoing merging enforces a central cusp, 
consequently we may have overestimated the mean value of $\kappa$ to some 
extent.

Our numerical results do not indicate any 
significant variation of the exponent $\kappa$ towards 
the center, see Fig.~\ref{fig-sig-phi}. Only in the case of the 
highest resolved halo Gal a trend of profile flattening can be noticed. 
Generally, a core with constant velocity 
dispersion might exist one a sub-resolution scale. 
The existence of such a core would weaken the restrictions for $\kappa$. 
However, the apparent inner constancy of $\kappa$ in the numerical simulations 
rather points to the existence of a power-law asymptote.

A further benefit of analyzing the VDPR is that this relation enables the 
determination of the leading processes responsible for the quite universal halo 
density 
profiles. The VDPR consists of two parts, the power-law $\Phi^\kappa$ 
and the 
relative potential  $\Phi_{\rm{out}} - \Phi$. The latter can be understood as 
the gain in energy 
of the particles falling into the potential well. Utilizing the virial 
theorem $2T(r) = U(r)$, where 
$T(r)  = (1/2)\sum\nolimits_{r<r_i<r+\Delta r}m_iv_i^2$ and 
$U(r)  =\sum\nolimits_{r<r_i<r+\Delta r}m_i(\Phi_{\rm{out}}-\Phi)$ 
and assuming isotropy for the velocity 
dispersion, $3\sigma_r^2 = v_i^2$ we obtain $\sigma_r^2 = 
(1/3)(\Phi_{\rm{out}}-\Phi)$. The factor 1/3 is 
very close to the value $a=0.29$. Dark matter 
halos, therefore, can be considered to be locally virialized in their outer 
region. This leads to the observed, steep outer profiles. In the inner region of 
the halo the additional factor $\Phi^\kappa$ occurs, which may be 
attributed to the motion of relatively slow particles concentrated within the 
central potential well.
Consequently, slow particles are more likely to be found in the center. This 
argument is scale-independent and should therefore result either in a power-law 
for 
$\sigma_r^2(\Phi)$ or in a dispersion core in the center 
of the halo.  The characteristic and universal density profiles are a result 
of these two mechanisms, which do neither depend on the history of the halo 
formation nor on the cosmological environment. For a perfectly relaxed 
system an universal density profile should be expected.

In summary, we have shown that the radial velocity dispersion can be 
approximated well by the VDPR as a function of the potential. We have provided 
physical arguments for the analytical form of the VDPR. The numerical data 
already obey the inner power-law over one order in magnitude. Considering the 
inner asymptotics, we have shown that the density profile becomes a power-law if  
the potential is below $\approx 0.001\times  \Phi_{\rm out}$. We conclude that 
the density profile can reach its asymptotic slope $n\lesssim 0.58$ only at 
scales smaller than those presently resolved.

\section{Acknowledgment}
We gratefully thank Anatoly Klypin, Gustavo Yepes and Matthias Steinmetz for 
providing
results of numerical simulations. Most of the simulations have been performed at 
the
Leibniz Rechenzentrum M\"unchen and at the Konrad-Zuse-Zentrum f\"ur
Informationstechnik Berlin.

\newcommand{\nucphys}{Nucl. Phys.}
\newcommand{\PRL}{Phys. Rev. Lett.}
\newcommand{\TUB}{Technische Universit{\"a}t Berlin}
\newcommand{\ZfPhys}{Z. Phys.}

\bibliography{ne-astro}
\bibliographystyle{apj}

\end{document}